\documentstyle[12pt,citesort]{article}
\def\IJMP #1 #2 #3 {{\it Int.\ J.\ Mod.\ Phys.}\ {\bf #1}\ (#2) #3}
\def\MPL #1 #2 #3 {{\it Mod.\ Phys.\ Lett.}\ {\bf #1}\ (#2) #3}
\def\NC #1 #2 #3 {{\it Nuovo Cim.}\ {\bf #1} (#2) #3}
\def\NP #1 #2 #3 {{\it Nucl.\ Phys.}\ {\bf #1}\ (#2) #3}
\def\PL #1 #2 #3 {{\it Phys.\ Lett.}\ {\bf #1}\ (#2) #3}
\def\PR #1 #2 #3 {{\it Phys.\ Rev.}\ {\bf #1}\ (#2) #3}
\def\PP #1 #2 #3 {{\it Phys.\ Rep.}\ {\bf #1}\ (#2) #3}
\def\PRL #1 #2 #3 {{\it Phys.\ Rev.\ Lett.}\ {\bf #1}\ (#2) #3}
\def\RMP #1 #2 #3 {{\it Rev.\ Mod.\ Phys.}\ {\bf #1}\ (#2) #3}
\def\ZP #1 #2 #3 {{\it Z.\ Phys.}\ {\bf #1}\ (#2) #3}

\date{}
\begin{document}
\begin{flushright}
hep-ph/9406395\\
June 1994
\end{flushright}

\vspace*{0.5in}
\begin{center}
{\bf 	FOUR WEAK BOSON PRODUCTION\\
	IN TEV PHOTON--PHOTON COLLISIONS\\
	AND HEAVY HIGGS SIGNAL}\footnote{Based on invited talk given at the
{\it ``Workshop on gamma--gamma colliders"}, March~28-31, 1994, Lawrence
Berkeley Laboratory}\\
[1ex]	G. Jikia\\
[1ex]	{\it Institute for High Energy Physics} \\
	{\it 142284, Protvino, Moscow region, Russia} 
\end{center}
\begin{abstract}
We study the signals for a heavy Higgs boson in the processes
$\gamma\gamma\to WWWW$, $\gamma\gamma\to WWZZ$ at the photon linear
collider. The results are based on the first complete tree level calculation
for these reactions. Using a forward ``spectator" $W$ tag, central
``spectator" $W$ veto to suppress backgrounds from transverse $W$, $Z$
production we show that the invariant mass spectrum of central $WW$, $ZZ$
pairs is sensitive to Higgs boson with a mass up to 1~TeV at a 2~TeV linear
collider.
\end{abstract}

\section{Introduction}

One of the most challenging puzzles of contemporary particle physics is
whether Nature indeed makes use of the Higgs mechanism of spontaneous
electroweak symmetry breaking. If Higgs boson will be found below 800~GeV or
so, this will be a proof of the so called weak scenario of the symmetry
breaking. Otherwise the scenario of the strongly interacting electroweak
sector (SEWS) will take place (for recent reviews see {\it e.g.}
\cite{SEWS}). The study of SEWS is one of the major motivations to built the
next generation of colliders.  While the potential of hadronic colliders
(see {\it e.g.} \cite{Gold} and references therein) as well as linear
$e^+e^-$ colliders \cite{linear} to explore SEWS was extensively studied,
much less was done for $\gamma\gamma$ colliders \cite{gg}, which are the
main subject of this Workshop.

The would be ``gold-plated" channel for Higgs boson production at
a $\gamma\gamma$ collider
\begin{equation}
\gamma\gamma\to H\to ZZ\to (q\bar q) (l^+l^-)
\end{equation}
was shown recently to be suffered from very large background from continuum
$ZZ$ pair production through $W$ boson loop for the Higgs mass above 350~GeV
\cite{aazz}. So this reaction, although very promising for the measurement of
the two-photon Higgs width for $M_H\leq 300-400$~GeV, provides very poor
possibilities for studying of a heavy Higgs and SEWS (see also
\cite{berger}).

Another very interesting potential application of photon-photon collisions
at a high energy linear collider proposed recently \cite{Brodsky} is $WW$
scattering, as illustrated in Fig. 1. In this process each photon is
resolved as a $WW$ pair. The interacting vector bosons can then scatter
pair-wise or annihilate; {\it e.g.} they can annihilate into a Higgs boson
decaying into a $WW$ or $ZZ$ pair. In principle, one can use these processes
\begin{equation}
\gamma\gamma\to W^+W^+W^-W^-,\quad W^+W^-ZZ
\label{WWWW}
\end{equation}
for studying SEWS. In fact this reaction at photon-photon collider is an
analog of the reaction $e^+e^-\to \nu\bar\nu WW(\nu\bar\nu ZZ)$ at linear
$e^+e^-$ collider.  Event rates for the reaction (\ref{WWWW}) were estimated
using effective $W$ approximation (EWA) \cite{Kingman} for several models of
SEWS and quite optimistic conclusions were given. However, EWA  has a
limited accuracy at energies of 1-2~TeV and, moreover, it does not permit to
calculate the effects of the tag and veto cuts used to isolate the Higgs
signal ({\it e.g.} K.~Cheung \cite{Kingman} had to use the exact calculation
for the reaction $\gamma\gamma\to WWH$ to estimate the efficiencies of these
cuts).

In this paper we present results of the exact standard model tree level
calculation for the reactions (\ref{WWWW}). In Section~2 we present cross
sections for different polarizations of $WWWW$, $WWZZ$ final states.  In
Section~3 we show that Higgs boson with a mass up to 700~GeV should be
relatively easily observed in photon-photon collisions at a 1.5~TeV linear
collider. Then we will concentrate on the heavy ($m_H=1$~TeV) standard model
Higgs boson case as a prototype for models of SEWS and will show that its
signal can be observed at a 2~TeV linear collider. We conclude with some
brief remarks in Section~4.

\section{Cross sections}

Total cross sections for the processes of four gauge boson production in
$\gamma\gamma$ collisions as a function of c.m. energy are shown in Fig.~2.
For comparison also are shown cross sections for other typical reactions in
$e^+e^-$ and $\gamma\gamma$ collisions (from \cite{Xsections}). One notes
that four weak boson $WWWW(WWZZ)$ production in $\gamma\gamma$ collisions is
larger than $\nu\bar\nu WW(\nu\bar\nu ZZ)$ production in $e^+e^-$ collisions
at the same energy.  Of course, this does not mean that the Higgs boson signal
is larger in photon-photon collisions, as both cross sections are dominated
by transverse gauge boson contributions.

Cross sections for different initial and final polarization states and two
values of the Higgs mass $m_H=100$~GeV and $m_H=\infty$ are shown in Fig.~3
for four $W$ boson production. One can see that the cross sections are
slightly larger for equal initial photon helicities. The dominating
contributions come from the production of all four gauge bosons being
transverse and one longitudinal and the other three transverse $W$'s. The
ratio of $TTTT/TTTL$ is about $(60\div 70)$\%. Such large fraction of $TTTL$
polarization state is due to production of soft $W_L$ and the analogous
effect was observed earlier for $\gamma\gamma\to WWZ$ reaction
\cite{Xsections}. The contributions from two longitudinal weak bosons $TTLL$
and $TLTL$ are about an order of magnitude smaller than that for $TTTT$
production. The yield of $TLLL$ and $LLLL$ final states is even smaller,
however one can see that for infinite Higgs boson mass their contribution
can be orders of magnitude larger than for a 100~GeV Higgs. It is this rise
of the cross section of longitudinal electro-weak boson interactions that
signals SEWS \cite{SEWS}. This effect is illustrated in Fig.~4 where cross
sections of $TTTT+TTTL$ as well as of final states containing at least two
longitudinal gauge bosons production are compared for $m_H=100$~GeV and
$m_H=\infty$. As usual, we can define the heavy Higgs boson signal to be the
difference between the cross section with a heavy Higgs boson and the result
with a light Higgs boson, {\it e.g.}
\begin{equation}
\sigma(\mbox{signal for } m_H=\infty) =
\sigma(m_H=\infty)-\sigma(m_H=100\mbox{~GeV}).
\end{equation}
Consequently, cross section for light Higgs boson represents the background.
From Fig.~4 one can conclude that signal-to-background ratio is about 10\%
for total cross sections.

\section{Signal of heavy Higgs boson at photon linear collider}

The scattering reaction (\ref{WWWW}) leads to two scattered $W$'s or $Z$'s
emerging at large transverse momentum in the final state accompanied by two
``spectator" $W$'s at low $p_\perp$ focussed along the beam direction. And
the heavy Higgs signal can be observed in the invariant mass spectrum of the
two hard scattered weak bosons. To select these $W$'s or $Z$'s we number all
the final gauge bosons according to their angles $\theta_i$ with beam
direction:
\begin{equation}
|\cos\theta_1|>|\cos\theta_2|>|\cos\theta_3|>|\cos\theta_4|.
\label{Ordering}
\end{equation}
We are interested in the mass spectrum of the ``central" pair $m(V_3V_4)$,
where $V$ denotes $W$ or $Z$ and we also assume that hadronic decay modes of
the central pair will be observed, {\it i.e.} we will not distinguish $W$'s
from $Z$'s. The crucial point to note is that in the framework of EWA the
initial $W_L$'s participating in the $W_LW_L$ scattering have a
$1/(p_\perp^2+M_W^2)^2$ distribution with respect to initial photons from
which they are produced. This is to be contrasted with a
$p_\perp^2/(p_\perp^2+M_W^2)^2$ distribution of the initiating $W_T$'s,
leading {\it e.g.} to $W_TW_T$ scattering. Analogous effect is known to take
place for $W$ distribution in quark -- (anti-) quark or $e^+e^-$ collisions
\cite{SEWS,Gold}. The softer $p_\perp$ distribution in the $W_LW_L$ case has
an important consequence: the spectator $W$'s tend to emerge with smaller
$p_\perp$ and correspondingly smaller angle than those associated with the
background processes of $W_TW_T$ or $W_TW_L$ scattering. Therefore we will
divide four final gauge bosons in two pairs of forward (backward) $V_1V_2$
and central $V_3V_4$ according to ordering (\ref{Ordering}) and will impose
different cuts on these pairs. We will veto hard forward (backward) $W$'s
$|\cos\theta_{1,2}|<z_f$ and will also require that $|\cos\theta_{3,4}|>z_c$
to enhance the signal/background ratio. Although it is the invariant mass
$m(V_3V_4)$ which we are interested in, to separate four gauge boson
production from backgrounds which come from $\gamma\gamma\to W^+W^-$ and
$\gamma\gamma\to W^+W^-Z$ reactions we will also tag forward (backward)
spectators $V_{1,2}$ in the region outside the dead cone along the beam
direction $|\cos\theta_{1\div 4}|<z_0$, where $z_0$ is determined by the
acceptance of experimental installation. The experimental signature is then
given by four hard jets from $WW(ZZ)$ decay in the central region with a
branching ratio of 50\% and jets or leptons in forward and backward regions
from the decay of spectator $W$'s. We have not modelled $W$, $Z$ decays, so
cuts will be imposed on momenta of vector bosons.

In Tables 1 and 2 we summarize cross sections and selection efficiencies for
a set of cuts for $WWWW$, $WWZZ$ production at photon-photon collider
realized at 1.5 and 2~TeV linear collider taking into account photon
luminosity spectrum \cite{gg}. Cross sections are quite large, for example
about ten thousand events of four weak boson production will be observed at
2~TeV linear collider with $\int{\cal L}dt = 200$~fb$^{-1}$.  It is
important to note that increasing dead cone from $5^\circ$ to $10^\circ$ we
are loosing a large fraction of the signal events.

In Figures~5, 6, we present the invariant mass distribution of the central
$WW$, $ZZ$ pairs summed over $WWWW$ and $WWZZ$ final states for two
different cuts and c.m. energies of 1.5 and 2~TeV. The number of events
corresponds to integrated luminosity of 200~fb$^{-1}$. First, at 1.5~TeV we
see clear peaks from the Higgs resonance at 500 and 700~GeV. Thus, in
principle one can observe a Higgs boson heavier than 350~GeV at the
photon-photon collider. However these events emerge from $WW$ fusion and
have nothing to do with the two-photon Higgs width measurements. Also, it is
hardly possible to push the observable Higgs mass well above 700~GeV at
1.5~TeV machine. Secondly, at a 2~TeV linear collider one can observe a
signal from 1~TeV Higgs boson. For a dead cone of $5^\circ$ one can see a
very distinctive enhancement around the invariant mass of 1~TeV. For
$z_0=\cos(10^\circ)$ the signal is still statistically significant, although
not so pronounced. For the infinitely heavy Higgs boson the invariant mass
spectrum is almost structureless at 2~TeV and more energy is needed to
observe the picture analogous to that in Fig.~6. Event rates as well as
signal/background ratio and the statistical significance corresponding to
Figs.~5, 6 are given in Table~3. Comparing this table with Tables~1, 2 we
see that while the signal contributes only about 10\% to the
total cross section for $m_H=1$~TeV, appropriate cuts permit to enhance the
signal-to-background ratio by an order of magnitude.

\section{Discussion}

The most important question is, certainly, comparison of the potential of
photon-photon collider with that of other machines. For hadronic and
$e^+e^-$ colliders much more detailed investigations were done including
decays of final $W$'s and $Z$'s and detector simulations
\cite{SEWS,Gold,linear}. For example, conclusion was done \cite{linear}
that the signal from 1~TeV Higgs boson was distinguishable from the case of
massless Higgs at the c.m. energy of $e^+e^-$ collider of 1.5~TeV and
integrated luminosity of 200~fb$^{-1}$. However, it was also found that the
integrated luminosity of 310~fb$^{-1}$ and 80~fb$^{-1}$ were needed to
discriminate the $m_H=\infty$ signal at $3\sigma$ level at 2~TeV and 3~TeV
linear collider, respectively. So, we can very roughly estimate that
potential of 2~TeV linear collider in photon-photon mode is the same as that
of 1.5~TeV $e^+e^-$ collider, provided that their luminosities are the same.
But what we would like to stress, is that because of different conditions at
the interaction point luminosity of high energy photon-photon collider has a
much less restrictive upper bound than that for $e^+e^-$ collider \cite{gg}.
And it is even stated that such a huge luminosity as
$10^{35-36}$~cm$^{-2}$~s$^{-1}$ is technically achievable in photon-photon
collisions \cite{gg}. Therefore, if such a luminosity will be really
achieved and if it will be possible to make experiments at such a huge
luminosity, photon-photon option will become very competitive with normal
$e^+e^-$ mode of linear collider.

\section*{Acknowledgements}

I am grateful to M.~Berger, F.~Boudjema, S.~Brodsky, K.~Cheung, I.~Ginzburg,
T.~Han, F.~Richard, V.~Serbo and V.~Telnov for helpful discussions. I am
indebted to G.~B\' elanger for making available the figures for
$\gamma\gamma$ and $e^+e^-$ processes. Special thanks to organizers of the
Workshop for financial help and to A.~Sessler and M.~Chanowitz for kind
hospitality. The attendance at the Workshop was supported, in part, by the
International Science Foundation travel grant.

\newpage
\section*{Figure captions}
\parindent=0pt
\parskip=\baselineskip

Fig.~1. $WW$ scattering at a photon-photon collider.

Fig.~2. Typical cross sections of electroweak reactions at $\gamma\gamma$ and
$e^+e^-$ modes of linear collider. For $t\bar t$ production the top mass
was set to 130~GeV. The other subscripts refer to the mass of the Higgs.
For reactions $e^+e^-\to WW\nu\bar\nu,ZZ\nu\bar\nu$ Higgs mass was set to
zero.

Fig.~3. Cross sections for different polarization states of initial and
final particles of the reaction $\gamma\gamma\to W^+W^+W^-W^-$ for
$m_H=100$~GeV and $m_H=\infty$ as a function of $\gamma\gamma$ c.m. energy.
$TTLL$ ($TLTL$) curves correspond to final states with $W_TW_T$ and $W_LW_L$
pairs of the same- (opposite-) charges.

Fig.~4. Comparison between the cross sections for $m_H=100$~GeV and
$m_H=\infty$ for equal and opposite helicities of the initial photons. For
the reaction $\gamma\gamma\to WWWW$ solid line is the total cross section;
dotted line is the $TTTT+TTTL$ cross section; dashed line is the sum of
cross sections with at least two longitudinal final $W$'s. For the reaction
$\gamma\gamma\to WWZZ$ corresponding cross sections are denoted by solid,
dotted and dash-dotted lines.

Fig.~5. Event rates for $WWWW+WWZZ$ production in $\gamma\gamma$ collisions
at 1.5~TeV linear collider for $m_H=100$ (shaded histogram), 500 and 700~GeV
(hatched histogram) for different cuts.
 
Fig.~6. Event rates for $WWWW+WWZZ$ production in $\gamma\gamma$ collisions
at 2~TeV linear collider for $m_H=100$ (shaded histogram) and 1000~GeV for
different cuts.
 
\newpage
\section*{Table captions}
\parindent=0pt
\parskip=\baselineskip
\begin{sloppypar}

Table 1: Cross sections (in fb) and selection efficiencies for
$\gamma\gamma\to W^+W^+W^-W^-$ and $\gamma\gamma\to W^+W^-ZZ$ for
$m_H=100$~GeV, 1~TeV and $\infty$ at $\sqrt{s_{e^+e^-}}=1.5$~TeV including
various cuts.  
1) no cut; 
2)~$|\cos\theta_{1,2}|>0.9,\, |\cos\theta_{3,4}|<0.7$;
3)~$|\cos\theta_{1\div 4}|<\cos(10^\circ),\, |\cos\theta_{1,2}|>0.9,\,
|\cos\theta_{3,4}|<0.7$; 
4)~$|\cos\theta_{1\div 4}|<\cos(5^\circ),\, |\cos\theta_{1,2}|>0.9,\,
|\cos\theta_{3,4}|<0.7$; 
5)~$|\cos\theta_{1\div 4}|<\cos(5^\circ)$;
6)~$|\cos\theta_{1\div 4}|<\cos(10^\circ)$.
\end{sloppypar}

Table 2: The same as Table 1 but for $\sqrt{s_{e^+e^-}}=2$~TeV and slightly
different cuts.
1) no cut; 
2)~$|\cos\theta_{1,2}|>0.95,\, |\cos\theta_{3,4}|<0.7$;
3)~$|\cos\theta_{1\div 4}|<\cos(10^\circ),\, |\cos\theta_{1,2}|>0.95,\,
|\cos\theta_{3,4}|<0.7$; 
4)~$|\cos\theta_{1\div 4}|<\cos(5^\circ),\, |\cos\theta_{1,2}|>0.95,\,
|\cos\theta_{3,4}|<0.7$; 
5)~$|\cos\theta_{1\div 4}|<\cos(5^\circ)$;
6)~$|\cos\theta_{1\div 4}|<\cos(10^\circ)$.

Table 3: Event rates for signal ($S$) and background ($B$)  summed over
$WWWW$ and $WWZZ$ final states as well as signal/background ratio and the
number of standard deviations for two values of the dead cone angle. Cuts
correspond to those of Figs.~5, 6. The value of integrated luminosity of
200~fb$^{-1}$ is assumed and branching ratio of 50\% for hadronic decays of
$WW$, $ZZ$ pairs is included. At $\sqrt{s}=1.5$~TeV we require that the
invariant mass $M_{34}$ of central pair lie in the interval
400~GeV$<M_{34}<$~600~GeV for $m_H=500$~GeV and 500~GeV$<M_{34}<$~800~GeV
for $m_H=700$~GeV. For $m_H=1$~TeV and $\sqrt{s}=2$~TeV
450~GeV$<M_{34}<$~1.1~TeV. 
\newpage

\begin{table}
\begin{tabular}{|c|c|c|c||c|c|c|} \hline\hline
\multicolumn{7}{|c|}{$\sqrt{s_{e^+e^-}} =1.5$ TeV} \\ \hline
\multicolumn{1}{|c|}{}
&\multicolumn{3}{|c||}{$\gamma\gamma\to W^+W^+W^-W^-$}
&\multicolumn{3}{|c|}{$\gamma\gamma\to W^+W^-ZZ$} \\ \hline\hline
 $m_H$ & 100 GeV & 1 TeV & $\infty$  
& 100 GeV & 1 TeV & $\infty$  \\ \hline  \hline 
1 &
 30.6 fb    & 32.7 fb   & 32.3 fb &
 6.85 fb   & 7.91 fb   & 7.58    fb
\\  \hline 
2 &
2.94 (10\%) &3.39 (10\%) &3.29 (10\%) &
.632 ( 9\%) &.870 (11\%) &.789 (10\%) 
\\  \hline 
 3 &
1.52 ( 5\%) &1.64 ( 5\%) &1.63 ( 5\%) &
.331 ( 5\%) &.404 ( 5\%) &.376 ( 5\%) 
\\  \hline 
 4 &
2.63 ( 9\%) &2.96 ( 9\%) &2.87 ( 9\%) &
.566 ( 8\%) &.750 ( 9\%) &.684 ( 9\%) 
\\  \hline 
 5 &
28.3 (93\%) &29.9 (91\%) &29.6 (92\%) &
6.46 (94\%) &7.31 (92\%) &7.02 (93\%) 
\\  \hline 
 6 &
20.8 (68\%) &21.5 (66\%) &21.3 (66\%) &
5.07 (74\%) &5.54 (70\%) &5.38 (71\%) 
\\ \hline \hline    
\end{tabular}
\caption{}
\end{table}

\begin{table}
\begin{tabular}{|c|c|c|c||c|c|c|} \hline\hline
\multicolumn{7}{|c|}{$\sqrt{s_{e^+e^-}} = 2$ TeV} \\ \hline
\multicolumn{1}{|c|}{}
&\multicolumn{3}{|c||}{$\gamma\gamma\to W^+W^+W^-W^-$}
&\multicolumn{3}{|c|}{$\gamma\gamma\to W^+W^-ZZ$} \\ \hline\hline
 $m_H$ & 100 GeV & 1 TeV & $\infty$  
& 100 GeV & 1 TeV & $\infty$  \\ \hline  \hline 
 1 &
 61.1 fb   & 68.1 fb   & 65.6    fb &
 14.3 fb   & 17.9  fb  & 16.1 fb   \\  \hline 
 2 &
3.10 ( 5\%) &4.01 ( 6\%) &3.67 ( 6\%) &
.701 ( 5\%) &1.31 ( 7\%) &1.01 ( 6\%) \\  \hline 
 3 &
.691 ( 1\%) &.805 ( 1\%) &.791 ( 1\%)  &
.161 ( 1\%) &.238 ( 1\%) &.187 ( 1\%) \\  \hline 
 4 &
2.46 ( 4\%) &2.99 ( 4\%) &2.74 ( 4\%) &
.554 ( 4\%) &.909 ( 5\%) &.731 ( 5\%) \\  \hline 
 5 &
53.7 (88\%) &58.4 (86\%) &56.5 (86\%) &
12.9 (90\%) &15.5 (86\%) &14.2 (88\%) \\  \hline 
 6 &
33.6 (55\%) &35.6 (52\%) &34.5 (53\%) &
8.95 (63\%) &10.1 (57\%) &9.52 (59\%) \\ \hline \hline    
\end{tabular}
\caption{}
\end{table}

\begin{table}
\begin{tabular}{|c|c|cccc||cccc|} \hline\hline
\multicolumn{2}{|c|}{}
&\multicolumn{4}{|c||}{$z_0=\cos(10^\circ)$}
&\multicolumn{4}{|c|}{$z_0=\cos(5^\circ)$}\\ \hline
$\sqrt{s_{e^+e^-}}$, TeV & $m_H$, GeV & $S$ & $B$ & $S/B$ & $S/\sqrt{B}$ 
& $S$ & $B$ & $S/B$ & $S/\sqrt{B}$  \\ \hline
1.5 & 500  & 84 & 34 & 2.5  & 14  & 218 & 56 & 3.9 & 29 \\
    & 700  & 24 & 23 & 1.0  & 5.0 &  53 & 37 & 1.4 & 8.7 \\ \hline
2   & 1000 & 14 & 21 & 0.67 & 3.0 &  74 & 59 & 1.3 & 9.6 \\ \hline\hline
\end{tabular}
\caption{}
\end{table}


\begin{thebibliography}{**}
\bibitem{SEWS}
M.S. Chanowitz, in Proc. of the {\it 2-nd KEK Topical Conference on $e^+e^-$
Collision Physics}, Tsukuba, Japan, November 26-29, 1991; K.-I.~Hikasa, in
{Physics and Experiments with Linear $e^+e^-$ Colliders}, Saariselk\" a,
Finland, 1992, Ed. R.~Orava {\it et al.}, World Scientific, p.  451; T.~Han,
in {Physics and Experiments with Linear $e^+e^-$ Colliders}, Waikoloa,
Hawaii, 1993, Ed. F.A.~Harris {\it et al.}, World Scientific, vol. I,
p.~270.
\bibitem{Gold}
J.~Bagger, V.~Barger, K.~Cheung, J.~Gunion, T.~Han, G.A.~Ladinsky,
R.~Rosenfeld, and C.P.~Yuan, \PR D49 1994 1246.
\bibitem{linear}
K.~Hagiwara, J.~Kanzaki, and H.~Murayama, KEK Report No. 91-4 (March 1991);
Y.~Kurihara, R.~Najima, \PL B301 1993 292; \
Y.~Kurihara, R.~Najima, KEK-Preprint-93-90, August 1993.
\bibitem{gg}
V.E.~Balakin, these proceedings; V.I.~Telnov, these proceedings and in
{Physics and Experiments with Linear $e^+e^-$ Colliders}, Waikoloa, Hawaii,
1993, Ed. F.A.~Harris {\it et al.}, World Scientific, vol. II, p.~551.
\bibitem{aazz}
G.V.~Jikia, \PL 298B 1993 224, \ \NP B405 1993 24; \
B.~Bajc, \PR D48 1993 1907; \
M.S.~Berger, \PR D48 1993 5121; \
D.A.~Dicus, C.~Kao, \PR D49 1994 1265; \
H.~Veltman, report SACLAY-SPHT-93-111, October 1993.
\bibitem{berger}
M.S.~Berger, these proceedings.
\bibitem{Brodsky}
S.~Brodsky, in {Physics and Experiments with Linear $e^+e^-$ Colliders},
Waikoloa, Hawaii, 1993, Ed. F.A.~Harris {\it et al.}, World Scientific,
vol.~I, p. 295.
\bibitem{Kingman}
K.~Cheung,
\PL B323 1994 85, \ and these proceedings.
\bibitem{Xsections}
M.~Baillargeon, F.~Boudjema, \PL B317 1993 371; \ M.~Baillargeon,
G.~Belanger, and F.~Boudjema, ENSLAPP-A-473-94, May 1994.

\end{thebibliography}
\end{document}